\documentclass{webofc}
\usepackage[varg]{txfonts}   
\usepackage{hyperref}
\usepackage{url}
\hypersetup{colorlinks=true,citecolor=blue,urlcolor=blue,linkcolor=blue}
\usepackage{slashed} 
\usepackage{caption,subcaption} 
\usepackage{nccmath} 
\usepackage{times}
\usepackage{amsmath} 
\usepackage{amssymb} 
\usepackage{amsopn} 
\usepackage{verbatim} 
\usepackage{graphicx} 
\usepackage{color} 
\usepackage{booktabs} 
\usepackage{empheq}
\usepackage{relsize}
\usepackage{fancyref}
\usepackage{float}
\usepackage{xcolor,colortbl}
\definecolor{olive}{rgb}{0.3, 0.4, .1}
\definecolor{fore}{RGB}{249,242,215}
\definecolor{back}{RGB}{51,51,51}
\definecolor{title}{RGB}{255,0,90}
\definecolor{dgreen}{rgb}{0.,0.6,0.}
\definecolor{gold}{rgb}{1.,0.84,0.}
\definecolor{JungleGreen}{cmyk}{0.99,0,0.52,0}
\definecolor{BlueGreen}{cmyk}{0.85,0,0.33,0}
\definecolor{RawSienna}{cmyk}{0,0.72,1,0.45}
\definecolor{Magenta}{cmyk}{0,1,0,0}
\definecolor{lcyan}{rgb}{0.6,1,1}
\usepackage[usenames,dvipsnames,pdf]{pstricks}
\usepackage{epsfig}
\usepackage{pst-grad} 
\usepackage{pst-plot} 
\usepackage{pstricks-add}
\usepackage[off]{auto-pst-pdf} 
 
\newcommand{\nths}{\negthickspace\negthickspace} 

\newcommand{\lp}{\left(} \newcommand{\rp}{\right)} 
  
\newcommand{\ls}{\left[}  \newcommand{\rs}{\right]}

\newcommand{\cd}{\!\cdot\!}
\newcommand{\st}[1]{\slashed{#1}}
\mathchardef\mhy="2D   

\DeclareMathOperator{\J}{J}

\RequirePackage{orcidlink}

\begin{document}
\title{Monte Carlo challenges for strong field quantum electrodynamics}
\author{\firstname{Anthony} \lastname{Hartin}\inst{1}\orcidlink{0000-0002-4881-8739}\fnsep\thanks{\email{A.Hartin@physik.uni-muenchen.de}}}
\institute{Ludwig Maximilian's University, Geschwister-Scholl-Platz 1, M\"unchen 80539, Germany}
\abstract{The assisted Schwinger effect, which is predicted to display non-perturbative quantum tunnelling, is expected to be produced in precision lab experiments with electron beams and intense lasers. Indeed, many novel effects predicted by a theory which incorporates the laser field exactly, can be probed experimentally with the proper application of theory and simulation. Among predicted first order effects are a rest mass shift and harmonic Compton scattering. Higher order effects include resonant transition rates which become apparent when scanned over specific kinematic parameters. The underpinning of the theory and experiment is a Monte Carlo simulation tool which precisely links experimental realities to the expected theoretical transition rates. We report on the progress of such a tool here.

} 
\maketitle


\section{Introduction}

The signature process of non-perturbative QED is the Schwinger effect, in which fermion pairs are produced from the vacuum by a background electromagnetic field. This can be understood as tunnelling through the quantum vacuum/background field at a rate that is exponentially suppressed \cite{Schwinger51,AiDrew20,GelTan2016}. \\

The Schwinger effect is central to many areas of physics, from the early universe to particle production in extreme conditions \cite{Hawking75,Tang2018,Baur07}. Though Schwinger pairs cannot yet be directly produced in a clean lab environment, a closely related process - assisted Schwinger production can. \\

In assisted Schwinger, or one photon pair production (OPPP), a single photon embedded in a background electromagnetic field converts to a pair through momentum contribution from the field. The transition rate of production shows a non-perturbative, exponentially suppressed region which is considered equivalent to that of Schwinger production from the vacuum. Indeed, a careful measurement of the rate would give an experimentally determined value for the Schwinger critical field itself \cite{Hartin19a}. \\

Assisted Schwinger production can be studied in purpose designed electron beam/intense laser experiments \cite{Bamber99,LUXEcdr21,facet100gev,eli16}. Moreover, the specific quantum field theory that determines its rate, also predicts the existence of other novel phenomena, such as harmonic scattering/production, mass shift and resonance production \cite{Hartin18a}. \\

Strong field quantum electrodynamics (SQED) is non-perturbative with respect to the intense background field, and perturbative with respect to bound fermion/gauge boson interactions. This means that it can be modelled by an event generator with suitable modifications. \\

A key ingredient of experiments designed to test SQED phenomena, is therefore a purpose built Monte Carlo program to generate strong field events. Such events depend, in part, on the beam and pulse characteristics of overlapping electron and laser beams \cite{Hartin18a}. \\

In this paper we will describe the challenges facing such a Monte Carlo program. This will necessitate an examination of the range of novel SQED phenomenology to reproduce, as well as the experimental schemas needed to generate this phenomenology. Finally, we will report on the progress of just such a Monte Carlo program, named {\bf IPstrong}.

\section{Strong field quantum electrodynamics (SQED)}

The quantum field theory used to predict SQED phenomena proceeds from the QED Lagrangian, which is organised so that an intense background electromagnetic field $A^e_\text{x}$ appears in the Dirac part of the Lagrangian. The effect of the background field is taken into account completely, via exact solutions of the Dirac particles embedded in that field. This is the so called Furry Picture \cite{Furry51}.

\begin{align}\label{eq:furpic}
\mathcal{L_{\text{QED}}^{\text{Furry}}}&=\bar\psi^\text{FP}(i\slashed{\partial}\!-\!e\slashed{A}^e_\text{x}\!-\!m)\psi^\text{FP}-\textstyle{\frac{1}{4}}(F_{\mu\nu})^2-e\bar\psi^\text{FP}\!\slashed{A}\,\psi^\text{FP} 
\end{align}

The Furry picture requires exact solutions of the Dirac equation in the background field, which for plane wave electromagnetic fields is known as the Volkov solution \cite{Volkov35}.

\begin{gather}
\psi_\text{prx}=V_\text{prx}\,e^{iS_\text{px}},\quad V_\text{prx}\equiv\ls 1-\mfrac{e\st{A}^e_\text{x}\st{k}}{2k\cd p}\rs\,u_\text{rp},\quad
S_\text{px}\equiv-p\cd x-\medint\int^{k\cdot x}\mfrac{2eA^\text{e}_\phi\cd p-e^2A^\text{e 2}_\phi}{2k\cd p}\,\text{d}\phi
\end{gather}

From this solution we get at least two insights. Firstly, the solutions of fermions in a background electromagnetic field describe a new particle - the bound fermion. The canonical momentum $\Pi_\text{px}$ of the bound fermion (equation \ref{eq:canmom}) has a dependency on space-time through its dependence on the background field potential and momentum. In other words, the bound fermion experiences a structured space-time which leads to dispersive effects.

\begin{gather}\label{eq:canmom}
\Pi_\text{px}=\partial_\text{x}\, S_\text{px}=p-eA^e_\text{x}+k\mfrac{2eA^e_\text{x}\cd p-e^2A^{e\, 2}_\text{x}}{2k\cd p}
\end{gather}

\begin{figure}[htb]
\centering\begin{subfigure}[t]{0.5\textwidth}
\centerline{\includegraphics[width=0.55\textwidth]{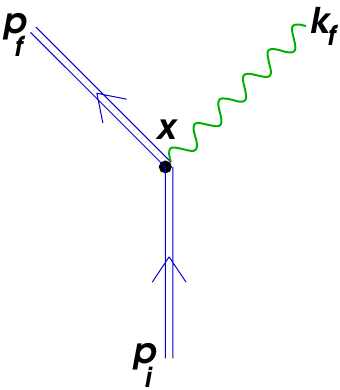}}
\caption{\bf One vertex HICS process.}\label{fig:hics}\vspace{0.1cm}
\end{subfigure}\begin{subfigure}[t]{.5\textwidth}
\centerline{\includegraphics[width=0.45\textwidth]{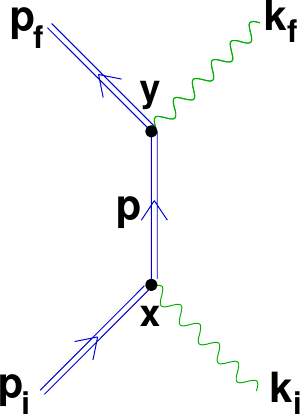}}
\caption{\bf Two vertex SCS process.}\label{fig:scs}
\end{subfigure}\vspace{-0.7cm}\caption{\bf Feynman diagrams for 1st and 2nd order strong field processes}
\vspace{-0.5cm}\end{figure}

Secondly, the theory is non-perturbative with respect to the background field but still perturbative as far as vacuum bosons $A$ and bound fermions $\psi_\text{prx}$ are concerned. So, we have a whole series of SQED processes to calculate (figures \ref{fig:hics} and \ref{fig:scs} show just two of the SQED processes that are possible). The Furry theorem (as opposed to the Furry picture), which states there can only be processes with an even number of vertices, is no longer operable. Now, odd vertex processes also exist with non zero transition probabilities. This is made possible because of momentum contributions from the background field. \\

The simplest processes in this theory are the one vertex processes, with one particle in the initial state. These processes describe decays of the bound fermions and SQED bosons. In the structured vacuum, the bound particles are unstable. The transition rates are well understood, having been analysed since the 1960s \cite{NikRit64a,NikRit64b,NarNikRit65,Hartin18a}. \\

Higher order SQED processes, which contain bound propagators, see a dispersive vacuum, which leads to an effective imaginary mass of the intermediate state. The inclusion of momentum from the background field allows the propagators $G^e_\text{yx}$ to go on shell. The result is that the higher order Furry picture transition rates display a series of Breit-Wigner type resonances \cite{Hartin06,Roshchup96,Bos79a,Oleinik67,Oleinik68}.

\begin{gather}
\quad G^e_\text{yx}=\int\mfrac{\text{d}p}{(2\pi)^4}E_\text{py}\;\mfrac{i(\st{p}+m)}{p^2-m^2+i\Gamma}\;\bar E_\text{px}
\end{gather}

\begin{figure}
\centering\begin{subfigure}[t]{0.5\textwidth}
\centerline{\includegraphics[width=0.95\textwidth]{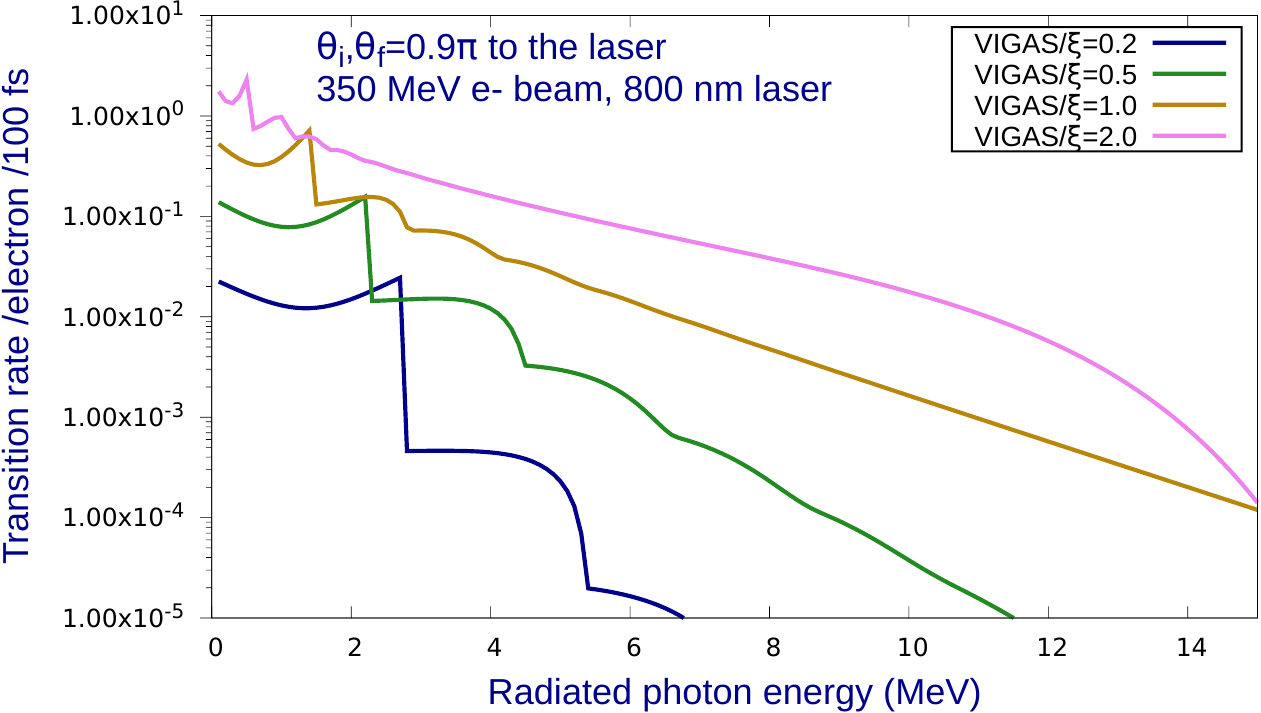}}
\caption{\bf Harmonic Compton edges in HICS.}\label{fig:cedge}\vspace{0.1cm}
\end{subfigure}\begin{subfigure}[t]{.5\textwidth}
\centerline{\includegraphics[width=0.95\textwidth]{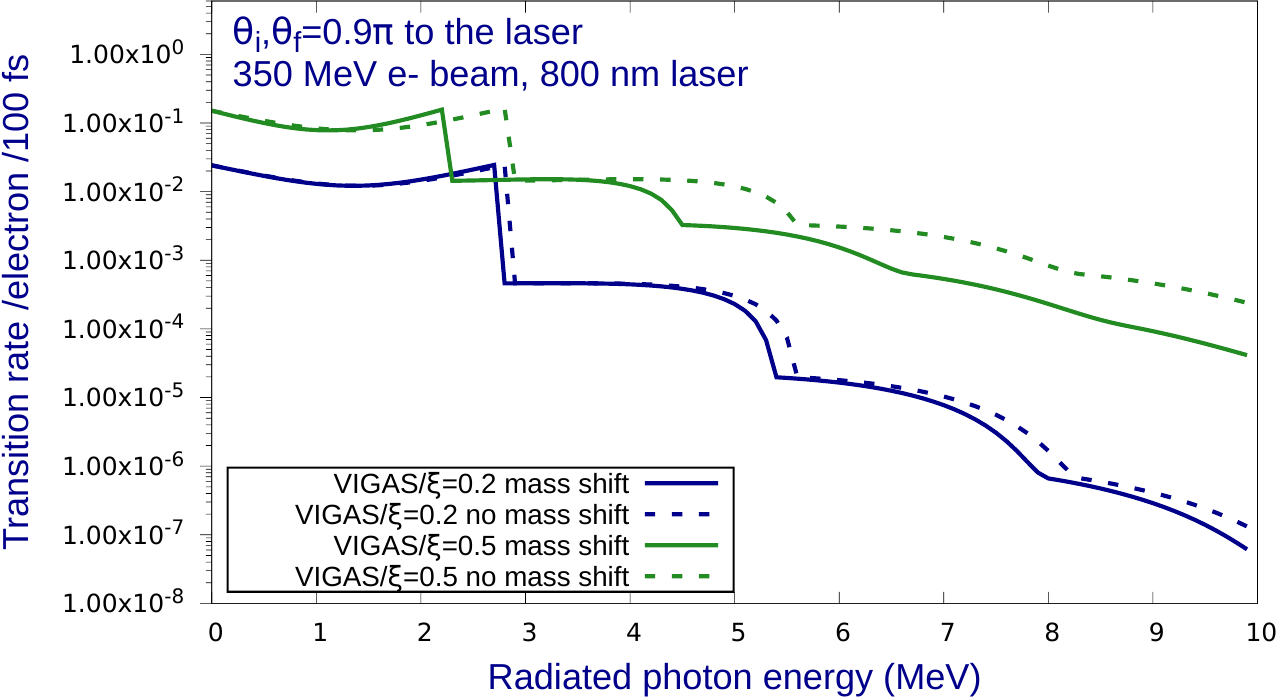}}
\caption{\bf Compton edge shift in HICS.}\label{fig:mshift}
\end{subfigure}\vspace{-0.7cm}\caption{\bf High intensity Compton scattering (HICS) phenomenology}
\vspace{-0.5cm}\end{figure}

As SQED theory proceeds to transition probabilities, several new phenomenological features are evident. Scattering involving harmonics of the background field, are displayed in a series of Compton edges in high intensity Compton scattering (HICS) (figure \ref{fig:cedge}). Moreover, each SQED Compton edge is shifted by an amount equivalent to a rest mass shift in the bound fermion compared to the free fermion (figure \ref{fig:mshift}). The extra contribution to the rest mass comes from the necessity of the bound fermion to exist with extra energy to account for an irreducible interaction with the background field. \\

\begin{figure}
\centering\begin{subfigure}[t]{0.5\textwidth}
\centerline{\includegraphics[width=0.95\textwidth]{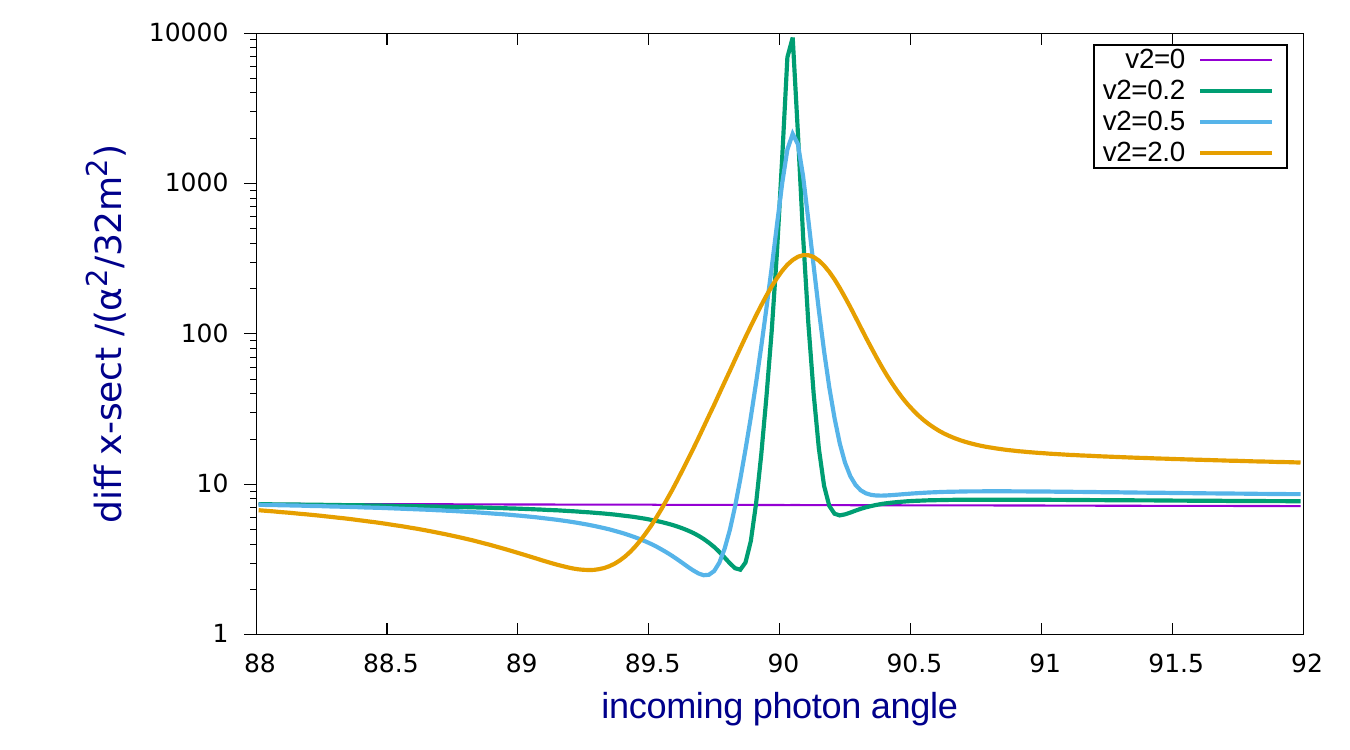}}
\caption{\bf Resonance with 2eV probe in SCS.}\label{fig:res1}\vspace{0.1cm}
\end{subfigure}\begin{subfigure}[t]{.5\textwidth}
\centerline{\includegraphics[width=0.95\textwidth]{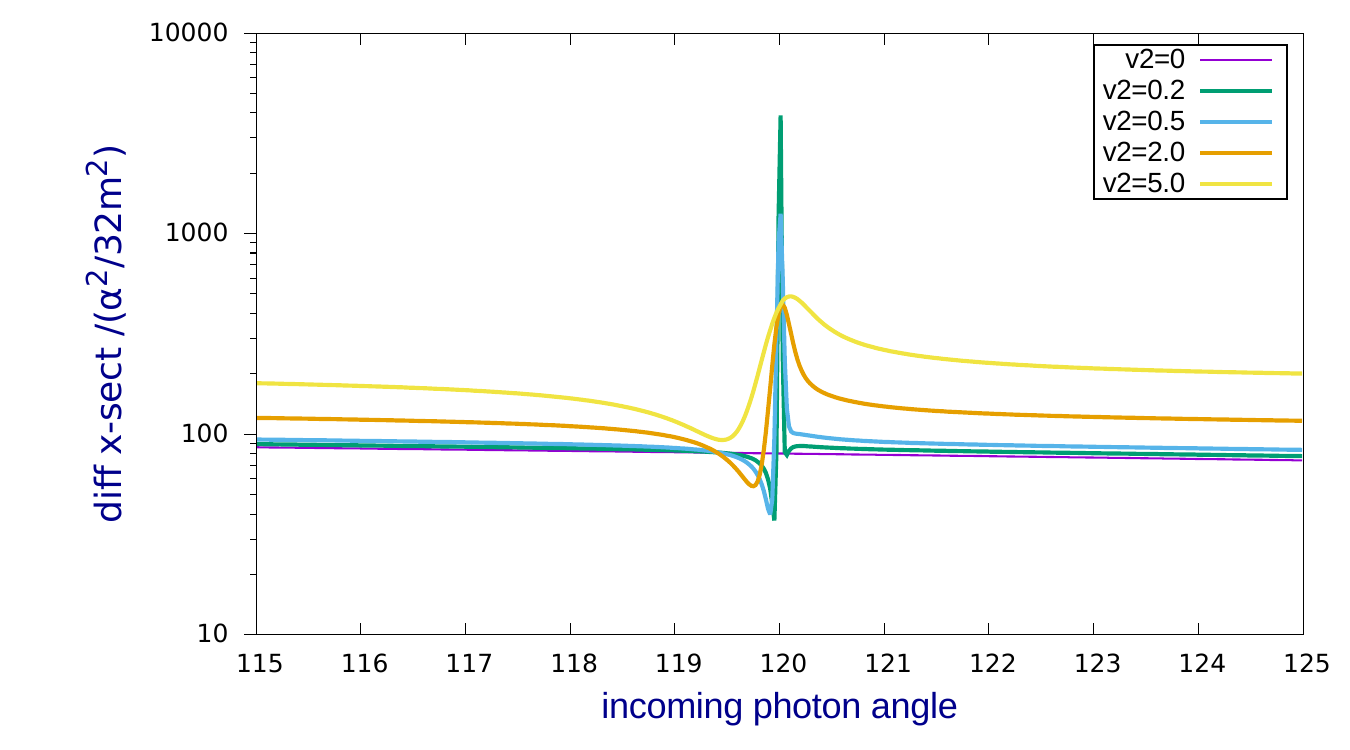}}
\caption{\bf Resonance with 4eV probe in SCS.}\label{fig:res2}
\end{subfigure}\vspace{-0.7cm}\caption{\bf Stimulated Compton scattering (SCS) phenomenology}
\vspace{-0.5cm}\end{figure}

The higher order SQED processes display resonant transition rates. As an example, the second order stimulated Compton scattering (SCS) display resonances in the differential rate when the kinematics of the process are scanned over (figures \ref{fig:res1} and \ref{fig:res2}). This is an indication of an energy level structure within the quantum vacuum which arises, phenomenologically, when the complete effect of background field is taken into account \cite{Zeldovich67}.

{\bf The first challenge for the Monte Carlo simulation of these effects, is theoretical. Only a few higher order SQED processes have been calculated. However, these higher orders display potentially dramatic phenomenology and must be included in simulation. SQED transition rate analytic forms are in general complicated, involving infinite summations over external field modes at each vertex. Efforts towards analytic simplification are ongoing.} \\

The experimental discovery of this SQED phenomenology, requires an analysis of the experimental setups designed to test them.

\section{SQED experimental schemas}

One of the most significant experiments designed to test SQED predictions, occurred at SLAC in the mid 1990s in the experiment labelled E144. In this experiment, the relativistic electrons from the SLAC linac were brought into near head on collision with a intense laser beam \cite{Bamber99} and specific SQED phenomenology was observed. A new series of planned experiments intends to use a similar setup with enhanced parameters that enable a new set of SQED experimental tests \cite{LUXEcdr21,facet100gev,eli16}. \\

\begin{figure}
\centering\begin{subfigure}[t]{0.6\textwidth}
\centerline{\includegraphics[width=0.5\textwidth]{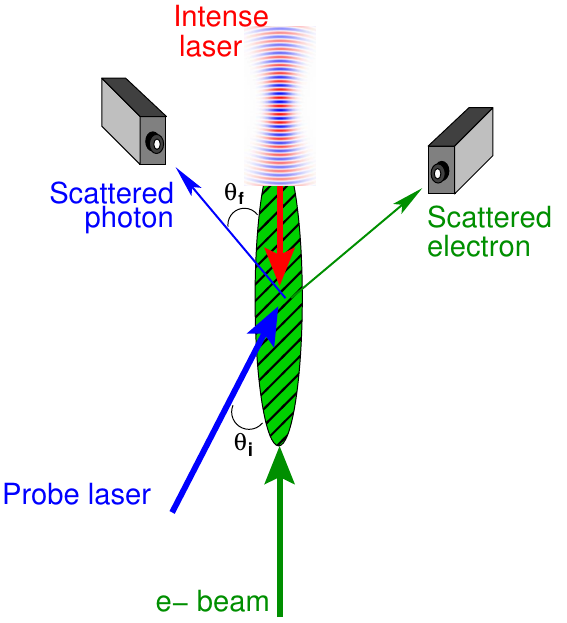}}
\caption{\bf Experimental schema for SQED tests.}\label{fig:schema}\vspace{0.1cm}
\end{subfigure}\begin{subfigure}[t]{.35\textwidth}
\centerline{\includegraphics[width=0.45\textwidth]{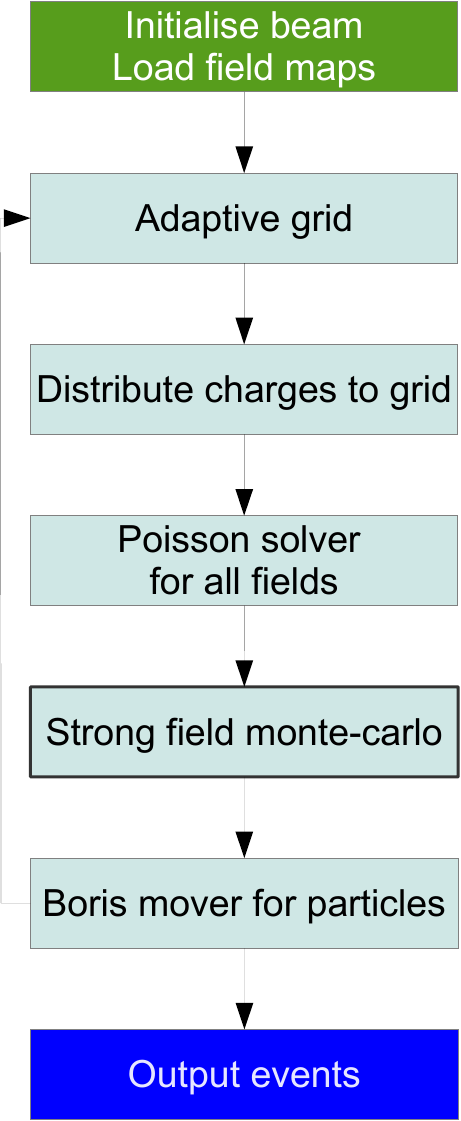}}
\caption{\bf IPstrong program blocks.}\label{fig:block}
\end{subfigure}\vspace{-0.7cm}\caption{\bf Experimental and program schemas.}
\vspace{-0.5cm}\end{figure}

An experimental schema that would test all of the phenomenology outlined in the previous section, would involve a similar setup as the SLAC experiment, with the addition of a probe laser to induce higher order Compton scattering and to scan over resonances (figure \ref{fig:schema}). These additional experimental components must be included in the IPstrong program modelling.\\

In order to accurately perform simulation, one has to take account of three SQED parameters. The first is the intensity parameter $\xi=e|\vec{A}^e_\text{x}|/m$ which quantifies the intensity of the laser pulse. Since a laser pulse is either Gaussian or flat top or some combination thereof, the intensity parameter varies throughout the laser pulse at each point of which, a SQED interaction takes place. \\

A second strong field parameter is the recoil parameter, which describes the strength of the background field in the rest frame of one of the participating particles, $\chi_e=\xi k\cd p/m^2$ for electrons and $\chi_\gamma=\xi k\cd k_i/m^2$ for the probe photons. To accurately simulate this parameter, one must take into account fluctuations within the electron beam, namely the energy spread and emittance. \\

The final strong field parameter is the resonance parameter, which is a combination of probe laser angle of incidence $\theta_i$ and photon energy $\omega_i$. To simulate this, we must take into account any energy jitter and accurately determine the overlap of three beams, that of the main and probe lasers with the electron bunch. \\

{\bf The second challenge for the Monte Carlo is the determination of the strong field parameters at each space-time point in the overlap of two or three interacting beams, taking into account energy and spatial fluctuations.} \\

Finally, we turn to the implementation of these requirements within a purpose built Monte Carlo program.

\section{Simulation of SQED processes}
\subsection{Overview}

The requirements for the simulation of the SQED phenomena outlined in the previous sections, necessitate a purpose built Monte Carlo program. No existing program provides simulation of SQED processes beyond first order. IPstrong specifically seeks to simulate the dramatic phenomenology of higher order resonant processes as well. \\

Since the beam fluctuations are to be accurately modelled, a combined particle-in-cell code is developed, which divides the 4D interaction space into 3D interaction voxels and a time step loop. The dimensions of the voxels are chosen small enough so that a constant laser intensity in that region is a reasonable approximation. \\

The intense laser that provides the background field is considered to be monochromatic at the nominal wavelength. This is reasonable because harmonic contributions are already taken into account within the SQED transition rates which sum over Fourier transformed modes of the laser. The probe laser is considered to be a source of monochromatic photons at the intensity indicated by its intensity pulse shape. \\

The incoming electron beam is modelled with energy spread and emittance using the principles of accelerator physics and beam optics. The combination of all factors leads to a precise determination of the parameters necessary to calculate the transition rate of each strong field process within each voxel at each time step. \\

Monte Carlo modelling of SQED transitions are performed within each voxel using rejection sampling, first to determine if a transition occurs and then to determine the final state parameters. If an incident particle does not undergo a quantum transition within a voxel, then it moves according to classical electrodynamics. \\

A program, named {\bf IPstrong}, has been developed that performs these steps to simulate first and second order SQED processes. {\bf IPstrong}'s structure bears similarities to previous SQED programs, CAIN \cite{Yokoya03} and Guinea-pig \cite{Schulte99}. However, {\bf IPstrong} goes beyond other programs in the range of strong field phenomena simulated. A top level view of the computing steps is shown in figure \ref{fig:block}.

\subsection{Algorithm}

\begin{figure}[h]
\vspace{-0.2cm}
\centerline{\includegraphics[width=0.95\textwidth]{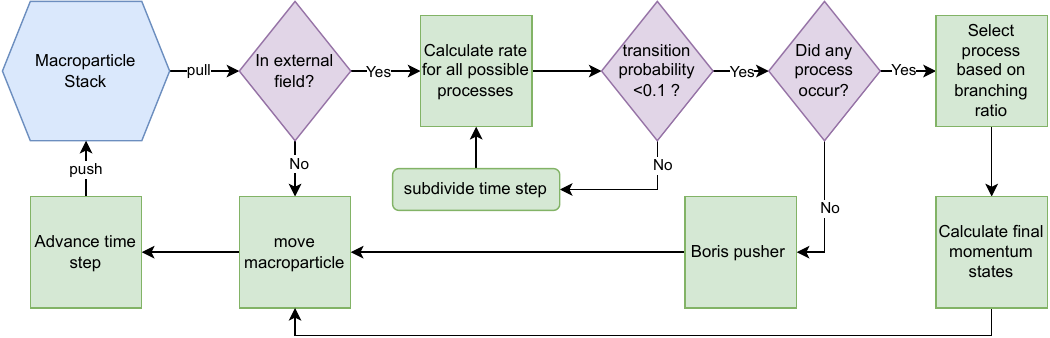}}
\caption{\bf IPstrong macroparticle processing}\label{fig:IPstrongflow}
\vspace{-0.5cm}\end{figure}

Figure \ref{fig:IPstrongflow} shows the inner logic of the program processing. Macroparticles are pulled from the stack at a specific time step, and the external (laser) field strength is calculated at the macroparticle's current position. With the initial state momentum and the field strength, the SQED transition rates are calculated. \\

Each transition rate calculation involves a numerical integration and an infinite summation over external field modes (see equation \ref{eq:hicsrate} as an example). Since this is too time consuming to calculate on the fly, lookup tables for the rates are provided externally via prior processing using Mathematica, and the precise rate is calculated via interpolation. Random throws establish whether the total rate is accepted and additional random throws pass the individual process according to the branching ratios.

\begin{gather}
\Gamma_{\!\text{HICS}}=-\mfrac{\alpha m^2}{\epsilon_\text{i}}\sum\limits_{n=1}^{\infty}\int_0^{u_n} \nths\mfrac{du}{(1+u)^2} \ls \J_{n}^2(z_u)-\mfrac{\xi^2}{4}\,\mfrac{1+(1+u)^2}{1+u} \lp\J_{n\text{+1}}^2(z_u)+\J_{n\text{-1}}^2(z_u)-2 \J_{n}^2(z_u)\rp\rs \notag\\
z_\text{u}\equiv \mfrac{\xi^2\sqrt{1+\xi^2}}{\chi_e}\ls u\lp u_n- u\rp\rs^{1/2},\quad u_n\equiv \mfrac{2\chi_e\,n}{\xi(1+\xi^2)} \label{eq:hicsrate}
\end{gather}

If the current time step is large enough that the total probability of a SQED transition exceeds 0.1, then the time step is subdivided in order to take into account possible multiple transitions. This is the same as the approach in \cite{Yokoya03}. The final state momentum is determined by energy-momentum conservation once sufficient momenta are determined by weighted random throws. \\

If a macroparticle fails to transition quantum mechanically in a time step, then it moves according to classical electrodynamics. For this purpose, a custom Lorentz invariant particle pusher, based on a Boris/leap frog scheme is implemented \cite{Vay07,Vay07a}. For accuracy, at least 100 time steps are required to span the wavelength of the external field.

\subsection{Validation}

\begin{figure}
\centering\begin{subfigure}[t]{0.5\textwidth}
\centerline{\includegraphics[width=0.95\textwidth]{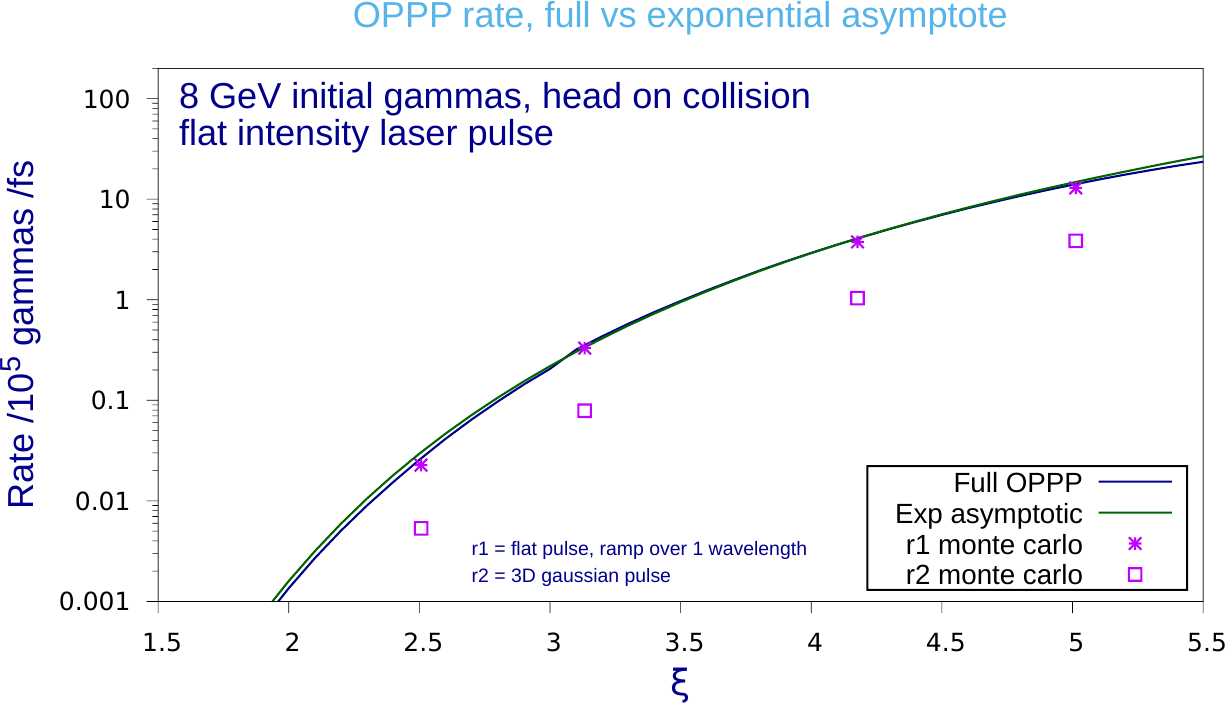}}
\caption{\bf Total rate of the OPPP process.}\label{fig:OPPPval}\vspace{0.1cm}
\end{subfigure}\begin{subfigure}[t]{.5\textwidth}
\centerline{\includegraphics[width=0.95\textwidth]{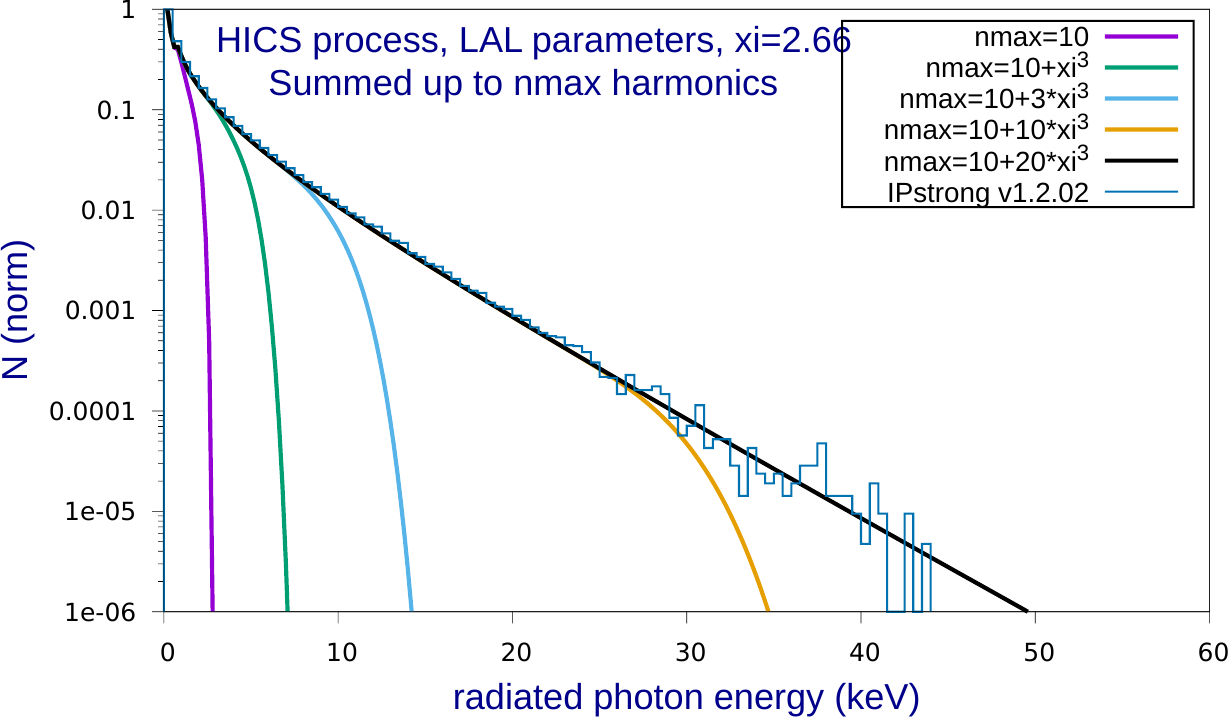}}
\caption{\bf Photon spectrum of the HICS process.}\label{fig:HICSval}
\end{subfigure}\vspace{-0.7cm}\caption{\bf IPstrong validation plots.}
\vspace{-0.5cm}\end{figure}

An ongoing process of validation is taking place, where {\bf IPstrong} results are benchmarked against analytic expectations. As an example, energetic photons in the initial state are brought into head on collision with an idealised flat intensity laser pulse. The total rate of the produced positrons is compared and shows reasonable agreement given the statistical significance. The extent to which the laser pulse varies from the flat ideal (i.e. an intensity ramp or a gaussian pulse) determines the closeness of the Monte Carlo points to the theoretical rate (figure \ref{fig:OPPPval}). \\

In a second validation example, the one vertex HICS process is summed up to very high harmonics of the background field. The {\bf IPstrong} Monte Carlo shows good agreement with produced photon spectrum over several orders of magnitude of the transition rate (figure \ref{fig:HICSval}). \\

{\bf The third challenge for the Monte Carlo is to continue the process of validation with regards to all strong field phenomenology modelled. In particular, these include the resonances of the higher order processes which are to be included and accurately reproduced.}

\section{Conclusion}

Strong field quantum electrodynamics entails unique non-perturbative processes such as Schwinger pair production, which may play an important role in the early universe, at the surface of black holes, in heavy ion collisions, future e+e- colliders and in electron beam/laser interactions. \\

Theoretical development of SQED is ongoing and directly impacts the design of purpose built Monte Carlo programs. Higher order processes with $n$ vertices, entail complicated analytic forms, with $2^n-1$ infinite summations over background field harmonics, as well as integrations over final state quantities. It is planned to generate transition rate tables of all relevant processes, in order to speed up Monte Carlo computation. \\

Experimental schemas to produce the phenomenology of all relevant SQED processes, entail the overlap of up to three beams. Real relativistic electron bunches are produced with emittance and energy spread. Laser pulses are usually Gaussian, though aggressive focussing to produce high intensities result in a combination of Gaussian and flat top pulse shapes. There is also associated jitter and various optical aberrations to take into account. \\

A specially built Monte Carlo program is necessary in order to precisely model the space-time overlap of two or three separate beams, taking into account experimental realities. A further requirement is the rigorous simulation of higher order SQED resonances. A particle in cell electromagnetic solver must be coupled with a Monte Carlo reproduction of non-perturbative transition rates. \\

Just such a purpose built non-perturbative Monte Carlo program, named {\bf IPstrong}, has been described here. The requirements have been specified and an ongoing process of validation has been outlined. This software tool is expected to be invaluable in planning and interpreting strong field, non-perturbative experiments in the coming years.

\end{document}